\pdfoutput=1
\documentclass[twocolumn]{article}
\usepackage{algorithm,algcompatible,amsmath,here,typearea,amsthm,color,cite,multicol,dblfloatfix,authblk,graphicx}
\typearea{2}

\newcommand{\ve}[1]{\mbox{\boldmath $#1$}}

\newcommand{\xraw}[1]{\ve{z}^{#1}}
\newcommand{\xfake}[1]{\ve{x}^{#1}_{\rm fake}}
\newcommand{\xdata}[1]{\ve{x}^{#1}_{\rm data}}
\newcommand{\xlabeled}[1]{\ve{x}^{#1}_{\rm L}}
\newcommand{\xunlabeled}[1]{\ve{x}^{#1}_{\rm UL}}

\newcommand{\LossG}[0]{L_{G}}

\newcommand{\expectation}[2]{}
\newcommand{\fig}[4]{
\begin{figure*}[h]
\centering
\includegraphics[#4]{#1}
\caption{#2}
\label{#3}
\end{figure*}
}
\newcommand{\fighalf}[4]{
\begin{figure}[h]
\centering
\includegraphics[#4]{#1}
\caption{#2}
\label{#3}
\end{figure}
}

\title{Quantum semi-supervised generative adversarial network 
for enhanced data classification}
\author{Kouhei Nakaji and Naoki Yamamoto}
\affil{Department of Applied Physics and Physico-Informatics \& Quantum Computing Center,
Keio University, Hiyoshi 3-14-1, Kohoku, Yokohama, 223-8522, Japan}
\date{\today}
\begin{document}
\maketitle

\begin{abstract}
In this paper, we propose the quantum semi-supervised generative adversarial network 
(qSGAN). 
The system is composed of a quantum generator and a classical discriminator/classifier 
(D/C). 
The goal is to train both the generator and the D/C, so that the latter may get a high 
classification accuracy for a given dataset. 
The generator needs neither any data loading nor to generate a pure quantum 
state, while it is expected to serve as a stronger adversary than a classical one thanks to its rich expressibility. 
These advantages are demonstrated in a numerical simulation. 
\end{abstract}


\section{Introduction}

We are witnessing active challenges to develop several type of enhanced machine learning schemes via quantum computing, i.e., the research field of quantum machine learning 
\cite{jacob16,siddarth18,maria16,vojtech18,carlo17,maria18,park20}. 
In particular, based on the possible higher expressibility of quantum circuits over the classical one, which has been proven theoretically and demonstrated experimentally 
\cite{bremner2011,bremner2016,bremner2017,farhi2016,arute2019}, the quantum machine 
learning is expected to provide alternative subroutines to improve the performance. 
For instance, the quantum classifier \cite{maria16,vojtech18,maria18,park20} is a quantum 
circuit that may be able to classify the input classical data with higher classification accuracy than 
conventional classical schemes such as the support vector machine. 
Note that many of those quantum machine learning applications need additional quantum operations or QRAM \cite{qram} to load the 
classical data into the quantum circuit, which eventually may vanish the possible quantum advantage.

The machine learning scheme focused in this paper is the generative adversarial network 
(GAN) \cite{gan}. 
In general, GAN consists of two adversarial components, typically a generator (generating 
a fake data) and a discriminator (discriminating a real or fake data), and their adversarial 
training yields an outperforming system over the one trained solely. 
GAN is also a topic actively studied in the quantum machine learning regime~\cite{qugan,seth18,haozhen18,qgan,marcello18,ling18,jinfeng18,jonathan19,shouvanik119,huang2020realizing,huang2020experimental,anand2020experimental,ahmed2020quantum,stein2020qugan,herr2020anomaly}. 
For example, Refs.~\cite{seth18,qugan} provide a method to train the quantum generative 
model by using GAN, where both the discriminator and the generator are quantum systems. 
Also \cite{haozhen18} provides a method to synthesize a quantum generative model for discrete 
dataset, in the GAN framework. 
A useful application of GAN was proposed in \cite{qgan}; their GAN consists of a quantum 
generator and a classical discriminator, and via the adversarial training the generator acquires 
the quantum state corresponding to a target probability distribution, which is then sent to 
another quantum circuit running a Grover-type algorithm (more precisely, the amplitude 
estimation algorithm for Monte Carlo simulation). 
Note that, however, this means that the generator is required to produce a target pure 
quantum state.

In this paper, we propose the {\it quantum semi-supervised GAN (qSGAN)}. 
As in the case of standard GAN, the SGAN 
\cite{semi-supervised-gan-1,semi-supervised-gan-2,semi-supervised-gan-3,semi-supervised-gan-review} 
consists of a generator and a discriminator, but the role of discriminator is to discriminate 
real/fake of the input data plus to estimate its label; that is, the goal of SGAN is to train the 
discriminator so that it becomes a good classifier, rather than realizing a high-quality generator. 
Our qSGAN consists of a quantum generator and a classical discriminator, like the case of 
\cite{qgan}, but the main goal is to train the classical discriminator rather than the quantum generator. 
In other words, the quantum system is used to train the classical system. 
Therefore, our qSGAN needs neither any data loading nor to generate a pure quantum state, while it is expected to serve as a stronger adversary than a classical one, based on its rich expressibility.

The rest of the paper is organized as follows. 
Section~\ref{section-Algorithm} is devoted to provide the algorithm of qSGAN. 
Section~\ref{section-Experiment} gives a numerical simulation to demonstrate the performance of qSGAN, with particular focus on the size of quantum circuit and the number of labeled training data. 
Also we show the expected noise-tolerant property of qSGAN. 
Moreover, we compare the quantum generator to a classical (deep) neural network generator and show that the resulting classification performance are comparable in a specific condition, suggesting a possible advantage of qSGAN for classification problems. 
Finally in Section~\ref{section-Conclusion}, we conclude the paper.


\section{Algorithm of qSGAN}
\label{section-Algorithm}

First let us recall the idea of standard GAN. 
GAN consists of a generator and a discriminator. 
The generator transforms a set of random seeds to samples (fake data). 
The discriminator receives either a real data from the data source or a fake data from the 
generator. 
Then the discriminator is trained so that it correctly classifies the received data into real 
or fake exclusively. 
Also the generator is trained so that its output (i.e., the fake data) are classified into real by 
the discriminator. 
Namely, the generator tries to fool the discriminator, and the discriminator tries to detect 
whether the received data is fake or real. 
If the training is successfully finished, then the generator acquires the ability to produce 
samples governed by a probability distribution that resembles the original distribution producing 
the real data.
In what follows, we describe the algorithm of qSGAN, based on the original proposal 
\cite{semi-supervised-gan-3}.

\begin{algorithm*}
  \caption{qSGAN}\label{algorithm-main}
  \begin{algorithmic}[1]
    \FOR{$j = 1$ to $N_{\rm iter}$} 
        \FOR{$a = 1$ to $N_B$}
            \STATE Load labeled data $\{(\xlabeled{a,1}, y^{a,1}), (\xlabeled{a,2}, y^{a,2}),\cdots, (\xlabeled{a,\ell}, y^{a,\ell})\}$ and 
            unlabeled data as $\{\xunlabeled{a,1}, \xunlabeled{a,2}, \cdots, \xunlabeled{a,m-\ell}\}$.
            \FOR{$i = 1$ to $m$}
                \STATE Set $\xfake{i}$ to the measurement result of $|\psi\rangle = U(\ve{\theta})|0\rangle$.
            \ENDFOR
            \FOR{$q = 1$ to $n$}
                \FOR{$i = 1$ to $m$}
  	            \STATE Set $\xfake{q(+)i}$ to the measurement result of $|\psi\rangle = U_{q +}(\ve{\theta})|0\rangle$.
  	            \STATE Set $\xfake{q(-)i}$ to the measurement result of $|\psi\rangle = U_{q -}(\ve{\theta})|0\rangle$.
  	            \ENDFOR
  	              	          	\STATE Set $\frac{\partial L_G}{\partial \theta_{q}}$ to $\frac{1}{2m}\sum_{i=1}^m \left[-\log D(\xfake{(q+)i}) + \log D(\xfake{(q-)i})\right]$.
  	        \ENDFOR
  	          	\STATE Update $\ve{\theta}$ by the gradient descent algorithm using $\frac{\partial L_G}{\partial \theta_{q}}$ ($q=1,2,\cdots, n$).
  	          	\STATE Set $L_D$ to $-\frac{1}{m} \sum_{i=1}^m\left[\log\left(1-D(\xfake{i})\right)+\log D(\xdata{a,i})\right]$.
  	          	\STATE Set $L_C$ to $\frac{1}{m}\sum_{i=1}^m h(c+1, C(\xfake{i}))+\frac{1}{\ell}\sum_{i=1}^{\ell} h(y^{a,i}, C(\xlabeled{a,i}))$.
  	            \STATE Set $L_{D/C}$ to $(L_D + L_C)/2$. 
  	            \STATE Compute the gradients of $C(\ve{x})$ and $D(\ve{x})$ by the back-propagation using $L_{D/C}$ and update the parameters by the gradient descent algorithm.
  	      \ENDFOR
  	     \ENDFOR
    \STATE The Classifier $C(\ve{x})$ as the final result
  \end{algorithmic}
\end{algorithm*}

{\bf (Source of the real data)} 
Suppose that we have $N_{\rm B}$ data batches, and each batch contains $\ell$ labeled 
and $m-\ell$ unlabeled data. 
We write the set of labeled data in the $a$-th batch as 
$\{(\xlabeled{a,1}, y^{a,1}), (\xlabeled{a,2}, y^{a,2}),\cdots, (\xlabeled{a,\ell}, y^{a,\ell})\}$ and 
that of unlabeled data as $\{\xunlabeled{a,1}, \xunlabeled{a,2}, \cdots, \xunlabeled{a, m-\ell}\}$, 
where $y^{a, i}$ is the label of the $i$-th labeled data in the $a$-th batch. 
We summarize $\xlabeled{}$ and $\xunlabeled{}$ into a single data vector as 
\begin{flalign}
    \xdata{a,i} = 
    \begin{cases}
    \xlabeled{a,i} &i\leq \ell\\
    \xunlabeled{a,i-\ell}&{\rm otherwise}
  \end{cases}.
\end{flalign}
The label data $y^{a,i}$ takes one of the values of $\{1, \cdots, c\}$, where $c$ is the 
number of classes.

{\bf (Generator)} 
The generator is a quantum circuit composed of qubits, which prepares the quantum state 
$|\psi\rangle=U(\ve{\theta})|0\rangle$, where $|0\rangle$ denotes the initial state of the circuit 
and $U(\ve{\theta})$ is the unitary matrix corresponding to the parametrized quantum 
circuit with parameters $\ve{\theta}$. 
The circuit outputs a fake data $\xfake{}$ as a result of the measurement of $|\psi\rangle$ 
in the computational basis. 
That is, $\xfake{}$ appears with probability 
${\bf P}(\xfake{})=\left|\left\langle \xfake{}\left| U(\ve{\theta})\right|0\right\rangle\right|^2$. 
Note that the stochasticity of the generator comes from this quantum intrinsic property, 
rather than the added random seeds like the classical GAN.

{\bf (Discriminator/Classifier)} 
In the SGAN framework, the discriminator $D(\ve{x})$, which judges real or fake for the 
received data $\ve{x}$, has an additional function, the label classifier $C(\ve{x})$. 
Hence, following Ref.~\cite{semi-supervised-gan-3}, we call this classical system simply 
the D/C. 
More precise description of those functions is as follows. 
First, $D(\ve{x})$ represents the likelihood that the received $\ve{x}$ came from the real data source. 
Next, $C(\ve{x})$ is a vector whose dimension is $c+1$; the $i$-th ($1\leq i \leq c$) component 
represents the likelihood that $\ve{x}$ belongs to $i$-th class, and the last one is the likelihood 
that $\ve{x}$ is a fake data. 
In this paper, we use a double-headed classical neural network to implement $C(\ve{x})$ and $D(\ve{x})$; 
\begin{flalign}
        C(\ve{x}) = g(f(\ve{x})), ~~~
        D(\ve{x}) = k(f(\ve{x})). 
\end{flalign}
$f$ is the function of the network shared by both $C(\ve{x})$ and $D(\ve{x})$, while $g$ and 
$k$ are the functions corresponding to the final layer; see Figure~\ref{neural-network}.

{\bf (Training rule)} 
In the training process in each batch, $m$ real data are loaded from the data source, and 
$m$ fake data are generated by the generator, which are sent to the D/C. 
Then the D/C assigns a label to those data via $C(\ve{x})$ and also judges real/fake via 
$D(\ve{x})$. 
Based on those results, the parameters of the generator and the D/C are updated, according 
to the following rule.

First, the generator is updated so that the generated fake data are classified into real by 
the discriminator. 
More specifically, we minimize the following cost function $L_G$ to update the parameters 
$\ve{\theta}$: 
\begin{flalign}
\label{L_G}
    \LossG &= {\bf E}_{\ve{x} \sim {\rm Generator}}(-\log D(\ve{x})) 
\nonumber\\
        &= \sum_{\ve{x}}\left(-\log D(\ve{x})\right)
                        \left|\left\langle \ve{x}\left| U(\ve{\theta})\right|0\right\rangle\right|^2
\nonumber\\
        &\simeq - \frac{1}{m} \sum_{i=1}^m \log\left(D(\xfake{i})\right), 
\end{flalign}
where $\xfake{i}$ is the $i$-th fake sample. 
Roughly speaking, minimizing $L_G$ corresponds to maximizing $D(\xfake{i})$, i.e., the 
likelihood that the fake data came from the real data source. 
In this paper, we take a quantum circuit where each parameter element $\theta_q$ is 
embedded into a single qubit gate in the form $U_q = \exp\left(- i \theta_q A_q /2 \right)$ 
with $A_q$ the single qubit operator satisfying $A_q^2={\bf 1}$. 
In this case, the gradient of $L_G$ is calculated as follows \cite{mmd}:
\begin{flalign}
       \frac{\partial L_G}{\partial \theta_{q}} 
           &= \frac{1}{2}\sum_{\ve{x}}\left(-\log D(\ve{x})\right)
             \left(\left|\left\langle \ve{x}\left| U_{q +}(\ve{\theta})\right|0\right\rangle \right|^2\right.
\nonumber\\&
     \quad \quad - \left.
        \left|\left\langle \ve{x}\left| U_{q -}(\ve{\theta})\right| 0 \right \rangle\right|^2\right),
\end{flalign}
where 
\begin{flalign}
     U_{q \pm}(\ve{\theta})
        &=U_{q \pm}(\{\theta_1, \cdots, \theta_{q-1}, \theta_q, \theta_{q+1}, \cdots, \theta_n\}) 
\nonumber \\ 
        &= U(\{\theta_1, \cdots, \theta_{q-1}, \theta_q\pm\pi/2, \theta_{q+1}, \cdots, \theta_n\}).
\end{flalign}
Thus, given the fake data $\xfake{(q\pm) i}$ obtained by measuring 
$U_{q \pm}(\ve{\theta})| 0 \rangle$, we get an unbiased estimator of 
$\partial L_G/\partial \theta_{q}$ as 
\begin{flalign}
       \frac{\partial L_G}{\partial \theta_{q}} \simeq 
           \frac{1}{2m}\sum_{i=1}^m \left[-\log D(\xfake{(q+)i}) + \log D(\xfake{(q-)i})\right].
\end{flalign}
This gradient descent vector is used to construct an optimizer for minimizing $L_G$.

Second, the D/C is trained so that it makes a correct real/fake judgement on the received data 
by $D(\ve{x})$ and, in addition, classifies them into the true class by $C(\ve{x})$. 
Hence, the parameters of the classical neural network are updated by minimizing the following cost 
function $L_{D/C}$:
\begin{flalign}
    L_{D/C} &= (L_{D} + L_{C})/2, 
\nonumber \\
    L_{D} 
       &= {\bf E}_{\ve{x} \sim {\rm Generator}}[-\log(1- D(\ve{x}))] 
\nonumber \\
         & \hspace{3em} + {\bf E}_{\ve{x} \sim {\rm Source}}[-\log(D(\ve{x}))] 
\nonumber \\
    & \simeq -\frac{1}{m} \sum_{i=1}^m\left(\log\left(1-D(\xfake{i})\right)+\log D(\xdata{a,i})\right),
\nonumber \\
    L_{C} &= {\bf E}_{\ve{x} \sim {\rm Generator}}[h(c+1, C(\ve{x}))] 
\nonumber \\
         & \hspace{3em} + {\bf E}_{\ve{x} \sim {\rm Labeled~Source}}[h(y^i, C(\ve{x}))]
\nonumber \\    
    &\simeq \frac{1}{m}\sum_{i=1}^m h(c+1, C(\xfake{i})) 
             + \frac{1}{\ell}\sum_{i=1}^{\ell} h(y^i, C(\xlabeled{a,i})).
\end{flalign}
Here $h(y, \xraw{})$ is the cross entropy of the discrete distribution $q(i) = \exp(z_i)/(\sum^{d}_{j=1}\exp(z_j))$ ($i=1,2,\cdots, d$) relative to the distribution $p(i) = \delta_{iy}$, defined as
\begin{flalign}
    h(y, \xraw{}) = - z_y + \log\left(\sum_{j=1}^{d}\exp(z_{j})\right),
\end{flalign}
where $z_j$ is the $j$-th element of the vector $\xraw{}$ and $d$ is the dimension of $\xraw{}$. 
As in the case of the generator, we use the gradient descent vector of $L_{D/C}$ to update the 
neural network parameters. 
In each iteration, the parameters are updated for all the batches. 
At the end of the training process with sufficiently large number of iterations, the trained D/C 
is obtained. 
The overall algorithm is summarized in Algorithm~\ref{algorithm-main} with $N_{\rm iter}$ as the number of iteration.


\section{Numerical demonstration}
\label{section-Experiment}

\fighalf{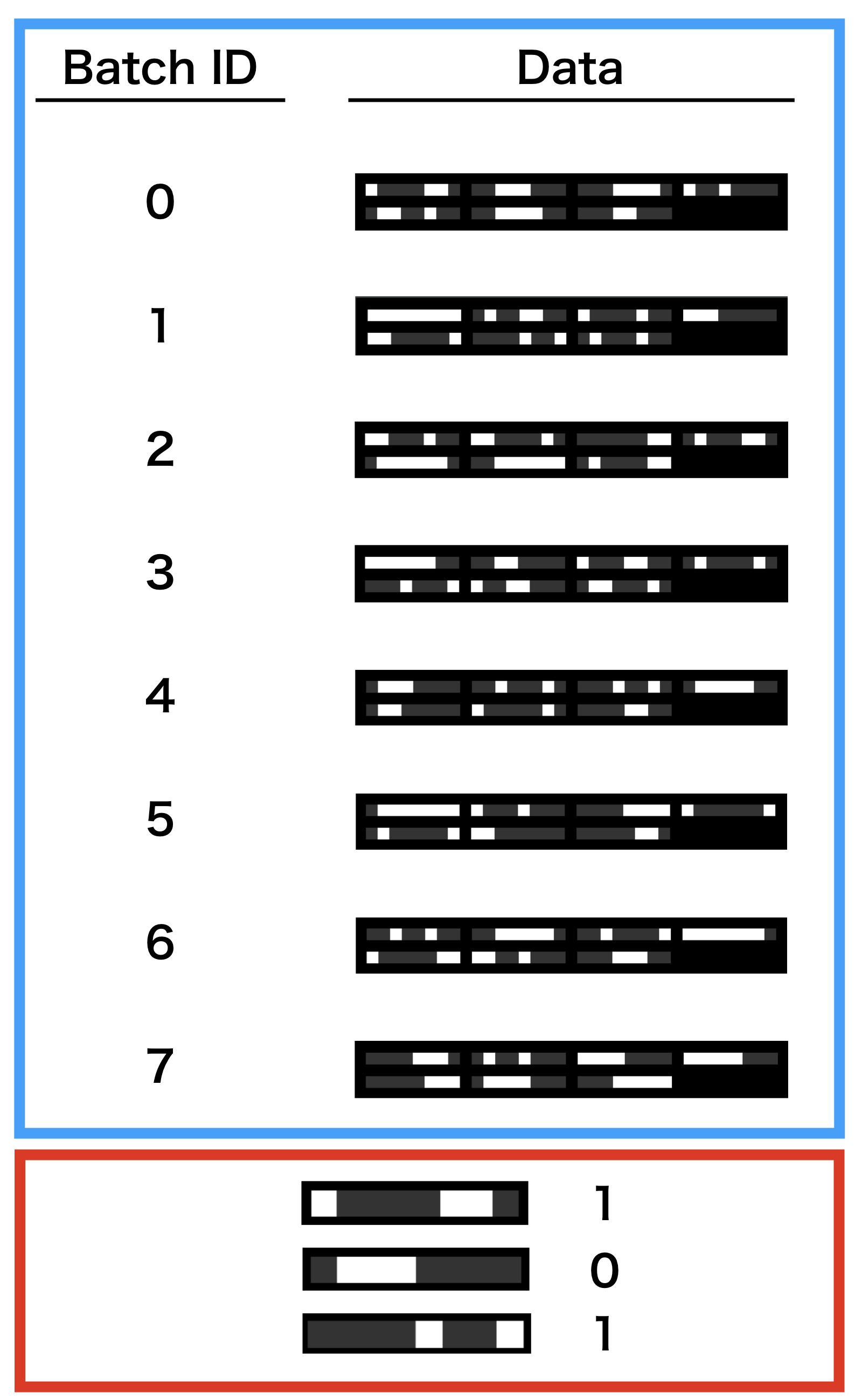}
{Top (enclosed by the blue rectangular): The dataset (= four batches) 
used in the numerical simulation. 
Bottom (enclosed by the red rectangular): Examples of images and 
their labels.}
{figure-dataset}
{width=200pt,bb=0 0 1600 2600}

In this section, we demonstrate the performance of the proposed qSGAN method by 
a numerical simulation. 
In particular, we will see that a quantum generator with higher expressibility leads to 
a better classifier, after successful training. 
Also the resulting classification accuracy is comparable to that achieved when using a standard classical neural network generator.

\subsubsection*{Problem setting and result}

The source of real data used in this simulation is a set of $1\times8$ pixel images, shown 
in Figure~\ref{figure-dataset}.
Each pixel takes the value of 0 (black) or 1 (white). 
Also the label `0' or `1' is assigned to each image (hence $c=2$) according to the following rule; 
if white pixels in an image are all connected or there is only one 
white pixel in an image, then that image is labeled as `0'; 
If white pixels in an image are separated into two disconnected parts, 
then the image is labeled as `1'. 
The number of images with label `0' and those with label `1' are both 28 (hence 56 images in total). 
The dataset is separated into eight batches, each containing $m=7$ images.

As the quantum generator, we use a 8-qubits parametrized quantum circuit with single layer or four layers; 
the case of four layers is shown in Figure~\ref{quantum-circuit}. 
Each layer is composed of parametrized single-qubit rotational gates 
$\exp(-i \theta_i \sigma_{a_i} /2)$ and CNOT gates that connect 
adjacent qubits; 
here $\theta_i$ is the $i$-th parameter and $\sigma_{a_i}$ is the Pauli 
operator ($a_i=x,y,z$). 
We randomly initialize all $\theta_i$ and $a_i$ at the beginning of 
each training. 
We run the numerical simulation on Qiskit QASM Simulator \cite{qiskit}.

\fighalf{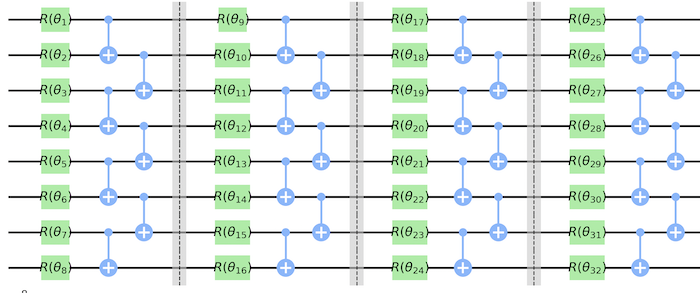}
{The quantum circuit with four layers used in the simulation.}
{quantum-circuit}{width=270pt,bb=0 0 380 147}

As the D/C, we use a neural network with four layers, shown in Figure~\ref{neural-network}. 
The first three layers are shared by both the discriminator $D(\ve{x})$ and the classifier 
$C(\ve{x})$. 
The number of nodes in the first, second, and third layer are 8, 56, and 8, respectively; 
all nodes between the layers are fully connected, and we use ReLU as the activation functions. 
The last layer for the classifier has three nodes, corresponding to the likelihood of label  ‘0’, label ‘1’, and fake classes; 
these nodes are fully connected to those of the third layer, and the softmax function is used 
as the activation function. 
The last layer of the discriminator has one node, simply giving the value of $D(\ve{x})$; 
this node is fully connected to the nodes of the third layer, and the sigmoid function 
($\sigma(x) = 1/(1 + \exp(-x))$) is used as the activation function. 
We implement the neural networks by PyTorch \cite{pytorch}.

\fighalf{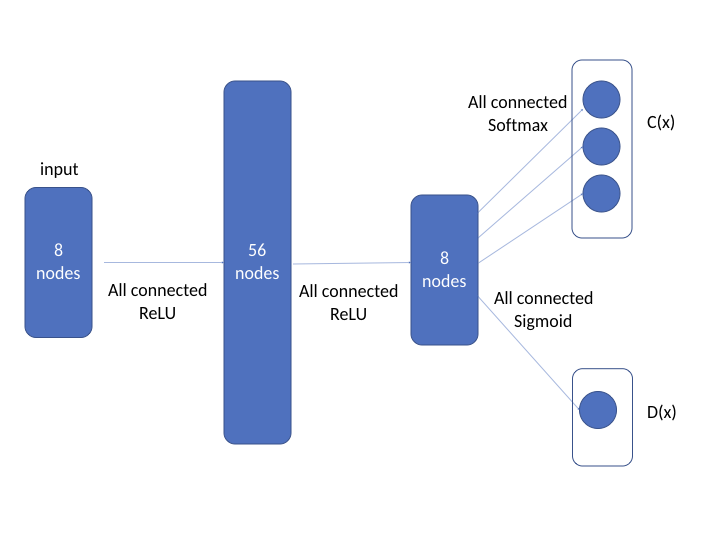}
{The D/C system, where the last layer functions as the classifier $C(x)$ 
or the discriminator $D(x)$.}
{neural-network}{width=350pt,bb=0 0 960 550}

In each trial of the algorithm, we choose two of the eight batches 
as the training dataset (hence $N_B=2$) and the other six as the test dataset. 
We perform the 4-fold cross validation by changing the training and test dataset, as summarized 
in Table \ref{table-datasest}. 
For each training/test dataset, we execute 20 trials (80 trials totally). 
To demonstrate the semi-supervised learning, some of the labels in each batch are masked; 
recall that the number of labeled example in each batch is denoted by $\ell$, which takes $\ell=2$ 
or $5$ in this simulation. 
As the gradient descent algorithm, we use Adam \cite{adam}, whose learning coefficient is set to 0.001 for the case of generator and 0.005 for the 
case of D/C.

\begin{table}[]
\begin{tabular}{|c|c|c|}
\hline
\begin{tabular}[c]{@{}c@{}}\textbf{Training Data}\\ (Batch ID)\end{tabular} & \begin{tabular}[c]{@{}c@{}}\textbf{Test Data}\\ (Batch ID)\end{tabular} & \textbf{\# of Trials} \\ \hline
0,1                    & 2,3,4,5,6,7        & 20                    \\ \hline
2,3                    & 0,1,4,5,6,7        & 20                    \\ \hline
4,5                    & 0,1,2,3,6,7        & 20                    \\ \hline
6,7                    & 0,1,2,3,4,5        & 20                    \\ \hline
\end{tabular}
\caption{The combination of the training/test dataset for the 4-fold cross validation. }
\label{table-datasest}
\end{table}

Figure~\ref{image-precision} shows the average classification accuracy for the test data versus 
the number of iteration, which are obtained as the average over 80 trials. 
The two subfigures are obtained with different number of labeled data, as $\ell=2$ and $\ell=5$. 
In each subfigure, three cases are shown, depending on the type of generator; 
the quantum generator with one layer (blue) and that with four layer (orange); 
also as a reference, the case of uniform-noise generator (green) that randomly generates 8-bit data 
with equal probability, which is not updated while training, is presented. 
The error bars represent the standard deviation of the average classification accuracy.

\fig{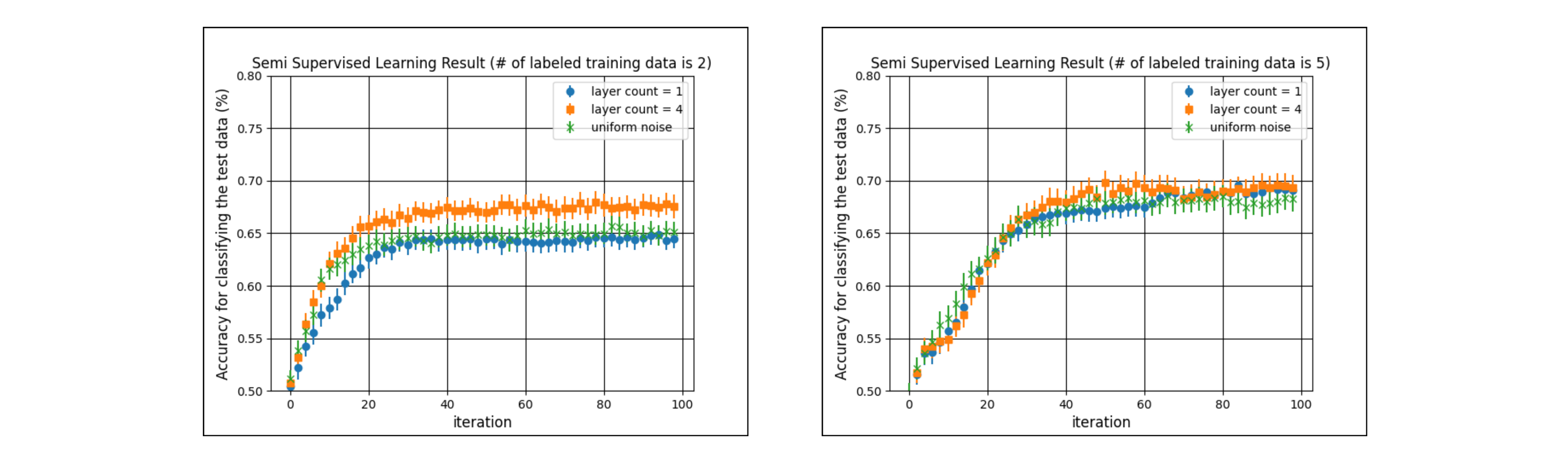}
{Classification accuracy of the classifier when using the quantum generator. 
The number of labeled data is $\ell=2$ (left) and $\ell=5$ (right).} 
{image-precision}{width=520pt,bb=0 0 2618 756}

We see that, when only a few labeled data is available ($\ell=2$), the quantum generator with four 
layers results in the highest classification accuracy, which implies that the quantum generator with 
bigger expressibility contributes to the higher accuracy by effectively generating samples to train 
the classical D/C. 
On the other hand, in the case where five of eight image data in each batch are labeled ($\ell=5$), 
the three generators achieve almost the same accuracy. 
This might be because, in this case, all the generators fail to generate more valuable dataset than 
the set of labeled real data, for effectively training the D/C. 
This observation is supported by the fact that the untrained uniform-noise generator, which of 
course is not related to the real dataset, achieves almost the same classification accuracy. 
Therefore, we expect that the quantum generator is useful when the number of labeled data is limited.

In this numerical simulation, we obtained the best classification accuracy when the constructed 
classical sample distribution corresponding to the output of the quantum generator does not 
match the distribution producing the real dataset, as predicted in \cite{dai2017good}. 
In addition, we found that the cost for the quantum generator, $L_G$, is larger than that for 
the classical D/C, $L_{D/C}$, when the best accuracy is reached. 
These facts are favorable for the current noisy quantum devices that cannot be effectively trained 
due to the noise. 
In the next subsection we will see how much the noise affects on the quantum generator and accordingly 
the classification accuracy.

\subsubsection*{Noisy qSGAN}

We examine the case where a noise channel is applied between the layers of the quantum generator. 
In particular we assume the depolarizing channel:
\begin{flalign}
\label{depolarization-noise}
    \mathcal{E}(\rho) = (1-p)\rho + p \frac{I}{2^n}, 
\end{flalign}
where $\rho$ is a density matrix, $I$ is the identity matrix, $n$ is the number of qubits, and $p$ 
is a noise parameter. 
Given the unitary operation $\mathcal{U}_i(\rho)=U_i \rho \, U_i^\dagger$ with the $i$-th depth 
unitary matrix $U_i$, the output density matrix is written as
\begin{flalign}
\label{total-noise}
     \rho_{\rm out} &=\mathcal{E} \circ \mathcal{U}_4 \circ \mathcal{E} \circ \mathcal{U}_3 
                    \circ \mathcal{E} \circ \mathcal{U}_2 \circ \mathcal{E} \circ \mathcal{U}_1
                         (|0\rangle\langle 0|)
\nonumber\\
         &=\mathcal{E} \circ \mathcal{U}_4 \circ \mathcal{E} \circ \mathcal{U}_3 
                    \circ \mathcal{E} 
                        \left( (1-p) U_2 U_1|0\rangle\langle 0|U_1^{\dagger}U_2^{\dagger} 
                            + p\frac{I}{2^n} \right)
\nonumber\\
         &= \cdots 
\nonumber\\
         &= (1-p)^4 \, U_4 U_3 U_2 U_1 |0\rangle\langle 0| 
                      U_1^{\dagger}U_2^{\dagger}U_3^{\dagger}U_4^{\dagger} 
\nonumber\\
         & \hspace{1cm} +\left(1-(1-p)^4\right)\frac{I}{2^n}.
\end{flalign}
The samples are generated by measuring $\rho_{\rm out}$. 
Other than the presence of noise, the simulation setting are the same as the noiseless case. 
We examine the case when only a few labeled data is available ($\ell=2$).

The resulting classification accuracy achieved via the quantum generator under the depolarization 
noise is shown in Figure~\ref{noisy-image-precision}, with several values of noise strength $p$. 
The horizontal axis represents the magnitude of total noise, i.e., the coefficient of the second term 
in Eq.~(\ref{total-noise}), while the vertical axis represents the average classification accuracy at the final 
(= 100-th) iteration step. 
As in Figure~\ref{image-precision}, the error bar is the standard deviation of the average accuracy. 
The result is that, as discussed at the bottom of previous subsection, the classification accuracy does not become worse than the noiseless case, as long as the depolarization noise for 
each layer of the quantum generator is suppressed to some level ($p=0.05$). 
This demonstrates the second advantage of the proposed qSGAN described in Section~1; 
that is, the quantum generator in our qSGAN framework does not need to generate a pure quantum state.

\fighalf{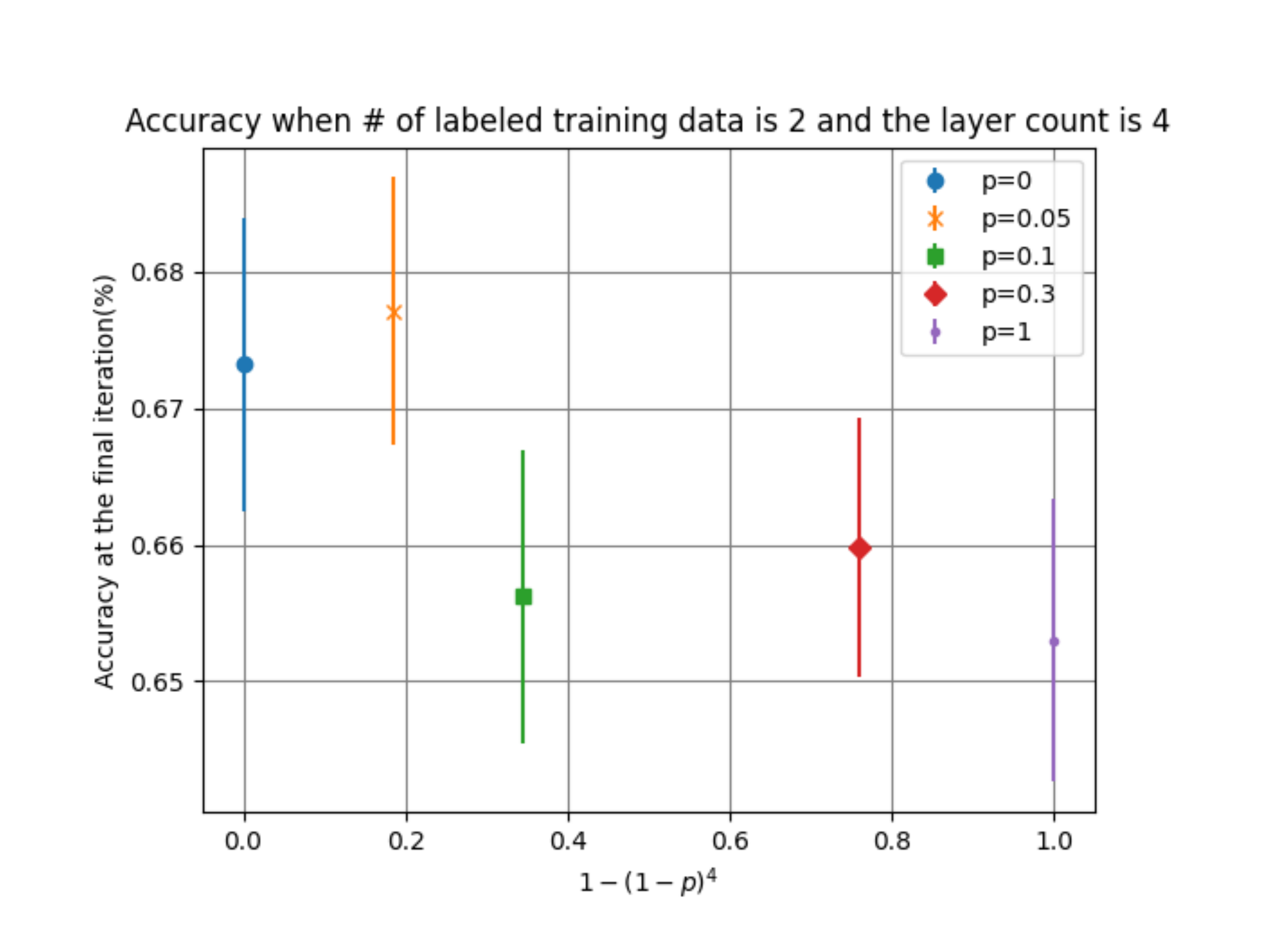}
{Classification accuracy of the classifier for $\ell = 2$ when using the four-layers quantum 
generator under the depolarization noise \eqref{depolarization-noise}. }
{noisy-image-precision}{width=350pt,bb=0 0 960 550}

\subsubsection*{Comparison with a classical neural network generator}

Finally, we compare the performance of the proposed qSGAN to the fully classical case where the 
generator is served by a five-layers classical neural network. 
The structure of this neural network is shown in Figure~\ref{classical-network}. 
The input to the network is the 1-dimensional normal Gaussian noise with zero mean and unit variance. 
The second, third, and fourth layer are composed of 40 nodes, and the fifth (= final) layer has 8 nodes. 
The nodes between the layers are fully connected and ReLU is used as the activation function. 
The output sample is obtained by transforming the values of the nodes at the final layer by the 
sigmoid function ($\sigma(x) = 1/(1 + e^{-x})$). 
We use the same D/C used in the quantum case. 
As the gradient descent algorithm, we use Adam, whose learning coefficients are set to 0.001 for 
both the generator and the D/C. 
Figure~\ref{image-precision-2} shows the classification accuracy for the test data over the number 
of iteration, which are obtained as the average over 80 trials when using the classical neural network 
generator. 
As in the case of Figure~\ref{image-precision}, the two subfigures are obtained with different number 
of labeled data, as $\ell=2$ and $\ell=5$.

The result is that, for the case $\ell=2$, the classifier aided by the classical neural network generator 
achieves the classification accuracy about 67$\%$, which is comparable to that of the four-layers 
quantum generator. 
The notable point is that the number of parameters of the classical and quantum generators are 
3688 and 32, respectively. 
Hence, naively, the quantum generator has a rich expressibility power comparable to the classical one 
even with much fewer parameters. 
This means that the training of the quantum generator is easier than the classical one, which is 
actually shown in Figures~\ref{image-precision} and \ref{image-precision-2}; 
about 40 iterations is enough to reach the accuracy 67$\%$ for the former case, while the latter 
requires roughly 400 iterations to reach the same accuracy. 
More importantly, this result implies that a bigger quantum generator with tractable number of 
parameters could have a potential to work even for some problems that are intractable via any 
classical one due to the explosion of the number of parameters.

\fig{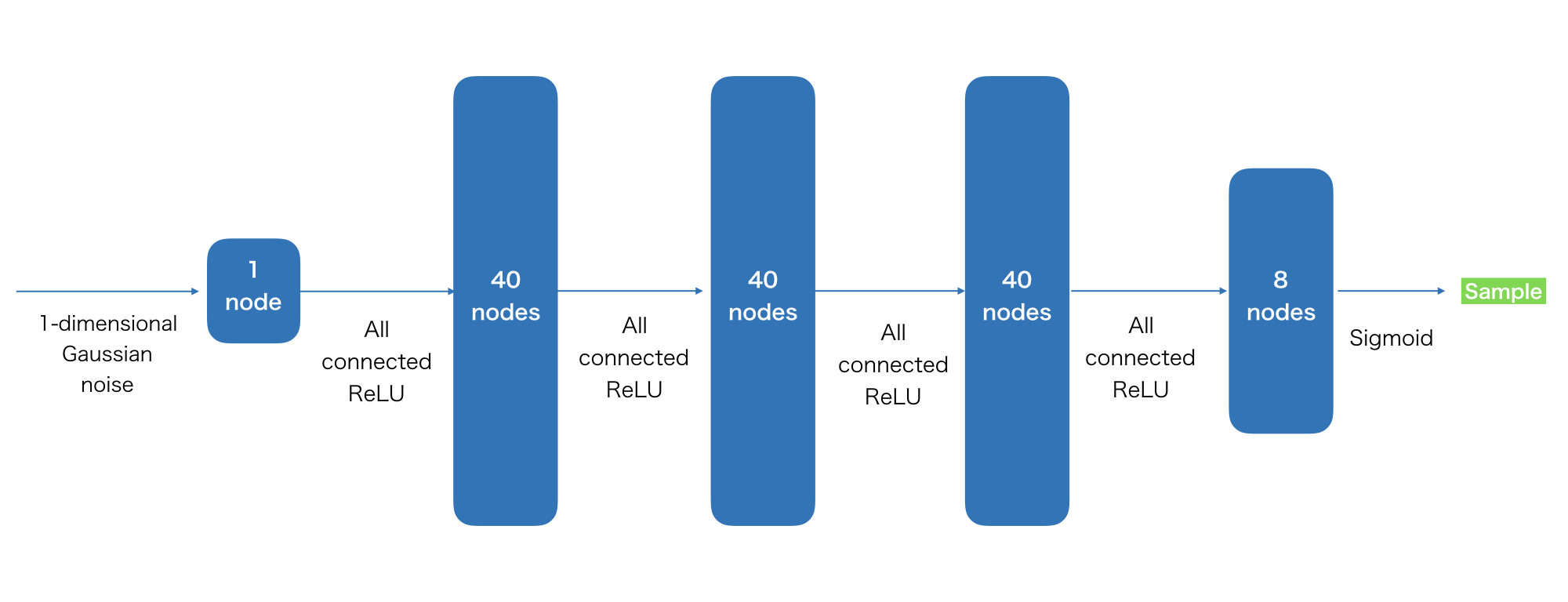}
{Structure of the classical neural network generator.}
{classical-network}{width=520pt,bb=0 0 2000 750}

\fig{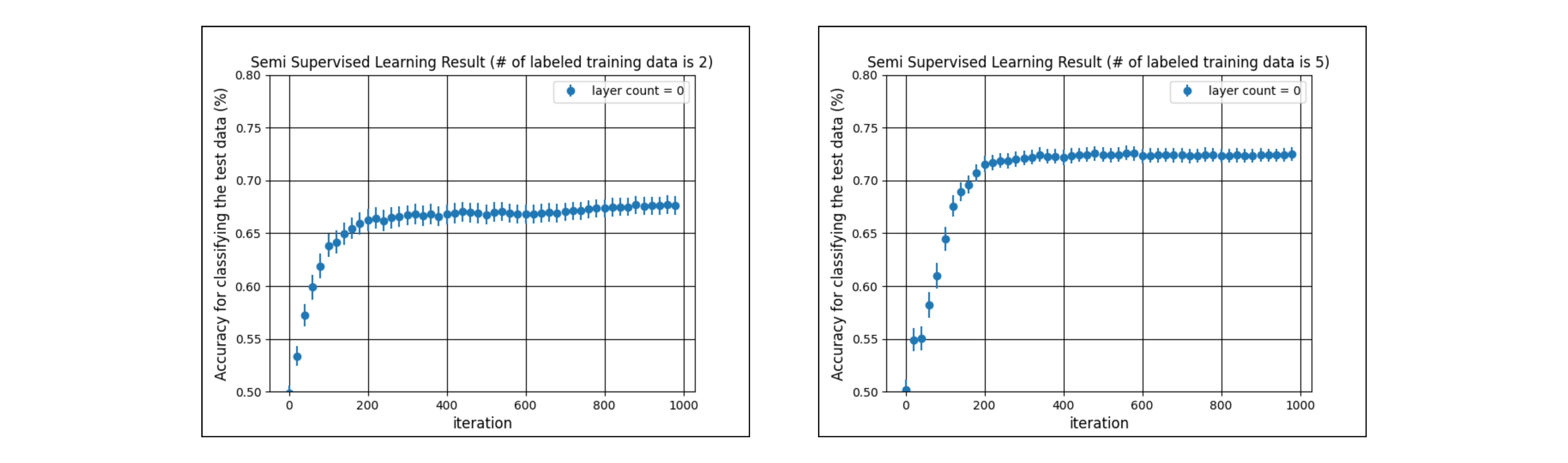}
{Classification accuracy of the classifier when using the five-layers classical neural network generator. 
The number of labeled data is $\ell=2$ (left) and $\ell=5$ (right).} 
{image-precision-2}{width=520pt,bb=0 0 2560 756}


\section{Conclusion}
\label{section-Conclusion}

In this paper, we provided qSGAN that performs a semi-supervised learning task 
by GAN composed of the quantum generator and the classical discriminator/classifier. 
This system has the following clear merits. 
That is, it needs neither data loading nor to generate a pure quantum state; rather its main role 
is to train the classical classifier, as a possibly stronger adversary than a classical one. 
The numerical experiment using the connected-disconnected image dataset shows that the rich 
expressibility of the quantum generator contributes to achieve the classification accuracy as high 
as that obtained when using the deep classical neural network generator (hence with much more 
parameters involved there). 
Also, we exemplified the noise-tolerant property of qSGAN under the depolarization noise, which is also a preferable feature for implementing  qSGAN on a current noisy quantum device.

\vspace{0.5cm}

This work is supported by the MEXT Quantum Leap Flagship Program 
Grant Number JPMXS0118067285 and JPMXS0120319794.
\bibliography{main} 
\bibliographystyle{unsrt}
\end{document}